\documentclass[aps,prl,twocolumn,groupedaddress,dvips]{revtex4}
\usepackage[dvips]{graphicx}
\begin{document}
\title{Mesoscopic Resistance Fluctuations in Cobalt Nanoparticles}
\author{Y. G. Wei, X. Y. Liu, L. Y. Zhang, and D. Davidovi\'c}
\affiliation{Georgia Institute of Technology, Atlanta, GA 30332}
\date{\today}
\begin{abstract}
We present measurements of mesoscopic resistance fluctuations in
cobalt nanoparticles and study how the fluctuations with bias
voltage, bias fingerprints, respond to magnetization reversal
processes. Bias fingerprints rearrange when domains are nucleated
or annihilated. The domain-wall causes an electron wavefunction
phase-shift of $\approx 5\pi$.  The phase-shift is not caused by
the Aharonov-Bohm effect; we explain how it arises from the
mistracking effect, where electron spins lag in orientation with
respect to the moments inside the domain-wall. Dephasing time in
Co at $0.03K$ is short, $\tau_\phi\sim ps$, which we attribute to
the strong magnetocrystalline anisotropy.
\end{abstract}

\maketitle

In micron scale metallic samples at low temperatures, interference
among scattered electron waves creates noticeable contributions to
sample resistance, including random but reproducible fluctuations
in conductance (CF).~\cite{washburn,lee2,altshuler4} One
remarkable consequence is that the resistance of phase-coherent
samples becomes sensitive to microscopic impurity configurations.
In this work we investigate the resistance of mesoscopic
ferromagnets at low temperatures and find a similar result that
the resistance is very sensitive to the magnetic state of the
sample. In particular, we observe significant wave-function
phase-shifts generated by domain-walls.

Mesoscopic effects in ferromagnets could be significantly
different from mesoscopic effects in normal
metals.~\cite{geller,loss1,tatara,adam1} While normal metals with
a short mean-free-path do not exhibit classical magnetoresistance
(MR), weakly disordered ferromagnets with a similar mean-free-path
display MR, which includes domain-wall resistance
(DWR)~\cite{kent,greg,viret,ebels,viret1,ruediger,kim,ravelosona,danneau,buntinx,yu1}
and anisotropic magnetoresistance (AMR)~\cite{mcguire}. MR could
lead to novel mesoscopic effects because the wavefunction-phase
depends on the scattering potential.~\cite{tatara,adam1}

\begin{figure}
\includegraphics[width=0.45\textwidth]{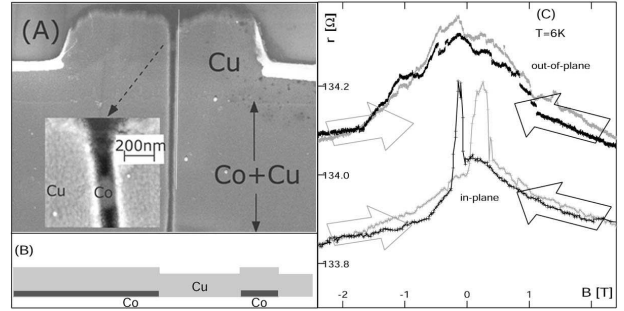}
\caption{A: Image of a typical sample. B: Sketch of the device
cross-section taken along the white line in Fig. 1-A. C:
Out-of-plane and in-plane MR.~\label{fig1}}
\end{figure}

Signatures of mesoscopic electron transport in ferromagnets have
been reported.~\cite{hong,aumentado,kasai,lees} However, the
dependence of the phase of the electron wavefunction on
magnetization reversal processes have not been measured yet. A
representative sample is shown in Fig. 1-A. A cobalt (Co) particle
of 200nm diameter and 10nm thickness is in electric contact with
two copper (Cu) leads of 50nm thickness. The gap between Cu leads
is 100nm. The device cross-section is displayed in
Fig.~\ref{fig1}-B.

Samples are made by electron beam lithography and shadow metal
deposition. First, a Co nanoparticle is deposited at $10^{-7}$
Torr base pressure and a rate of $0.5$nm/s. Deposition is stopped
when the Co thickness ($t$) reaches 10nm. Then we rotate the
sample without breaking the vacuum and deposit Cu at 0.5nm/sec.
The interface between Co and Cu is nearly free from adsorbates
because the Co surface is exposed to base pressure for less than
10 seconds. The nanoparticle is completely isolated from any other
ferromagnet in a vicinity of 1.5$\mu$m to remove the influence of
stray magnetic field from other magnetic parts of the device.

The sample is next exposed to air and transferred to a dilution
refrigerator. Surface oxidation reduces the metal thickness and
covers the film with a cobalt-oxide (CoO) layer. The time of air
exposure before evacuation in the dilution refrigerator is less
than one hour. We measure the resistance of Co films with various
thicknesses {\it in situ} and monitor how the resistance increases
when the films are exposed to air for one hour. From this we infer
that the thickness of Co-metal is reduced by approximately 2nm
after one hour of air exposure.  The sheet resistance of the Co
film at 4.2K temperature is $R_S\approx 120\Omega$.

CoO is antiferromagnetic and the Co/CoO interface generates an
exchange-bias effect in Co,~\cite{meiklejohn,hausmanns,buntinx}
which leads to pinning of the magnetization and the enhancement of
the coercive field. We expect that domain-walls are nucleated at
the interface between the exposed Co and unexposed Co (under Cu)
by applying well defined magnetic fields, analogous to the similar
behavior shown in Ref.~\cite{buntinx}. The nanoparticle can
support domains because the domain-wall width ($\delta_w$) in Co
($\delta_w=15nm$~\cite{viret,ebels,greg}) is much smaller than the
diameter.

Four samples were studied at low temperatures, and the
reproducible main result is that the correlation field is strongly
suppressed near zero field.  The dephasing time is also
reproducible among samples.

 Differential resistance ($r=dV/dI$) is
measured  as a function of DC-bias voltage ($V$) and the applied
magnetic field (B), ($r(V,B)$). The applied current is
$I+icos(2\pi ft)$, where $i=0.5\mu A$, and $f=80Hz$. Then, $r$ is
obtained by measuring the AC-voltage across the sample with a
lock-in amplifier.  At $0.03K$, we confirm that $r(V,B)$ is
independent of $i$ when $i<0.5\mu A$.
The resistance of Cu
leads is about $10\Omega$ and is not subtracted from $r$.

The out-of-plane (OP) and the in-plane (IP, magnetic field
perpendicular to the current) MR at $T=6K$ are shown in
Fig.~\ref{fig1}-B. First we discuss the IP-MR data. The magnetic
field is initially set to -12T, then it is reduced to -2.4T, and
then it is cycled between -2.4T and 2.4T.

The IP-MR graph has hysteresis. There are two sharp resistance
transitions in each field direction.  The low and the high field
transitions indicate nucleation and annihilation of domains,
respectively. The smaller coercive fields are symmetric, $B_C=\pm
62mT$, which can be explained if the magnetization is
 first reversed in unexposed Co (under Cu), as expected. The larger coercive
fields are $330mT$ and $-220mT$. The magnetic moments in exchange
biased Co change direction at these coercive fields. The coercive
field is larger in magnitude when $B$ increases, because $B$ is
initially $-12T$.

The resistance increases when the domains are nucleated. The
increase is explained by the AMR inside the wall, which arises
from the dependence of the conduction electron scattering rate on
the angle between the current and the
magnetization.~\cite{mcguire}

Next we discuss the OP-MR in Fig.~\ref{fig1}-B. OP-MR exhibits a
broad maximum at $B=0$ and a weak hysteresis. The maximum is
explained as arising from the rotation of magnetic moments
supported by the shape anisotropy. As the magnetic moments rotate
into the film plane, the angle between magnetic moments and the
current is reduced, so $r$ increases.

{\it Extracting CF's:} We describe the analysis of the OP-field
data. The analysis of the IP-field data is equivalent. The
resistance increases by $\approx 6\Omega$ when the temperature
drops from $6K$ to $0.03K$. Similar effect is reported in Co films
at temperatures above $1.5K$~\cite{brands} and is attributed to
enhancement of electron-electron interactions caused by
phase-coherence.~\cite{altshuler}

\begin{figure}
\includegraphics[width=0.45\textwidth]{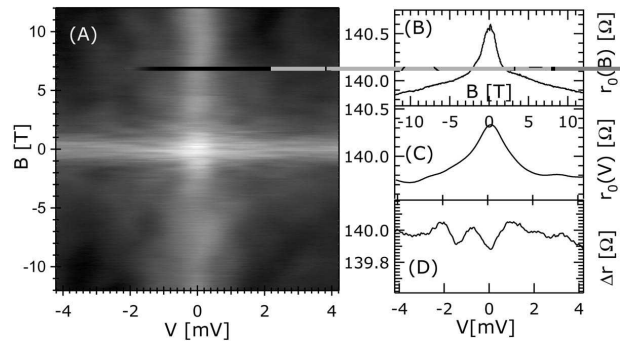}
\caption{A: Differential resistance ($r$) versus bias voltage and
out-of-plane magnetic field at 0.03K. B and C: Average resistance
versus out-of-plane field and bias voltage, respectively, defined
in text. D: Fluctuations in resistance with bias voltage,
$r(V,B)-r_0(V)+r_0$, at $B=-5.5T$.~\label{fig2}}
\end{figure}

To study mesoscopic effects, we obtain the dependence of $r$ on
two independent parameters, $V$ and $B$ ($r(V,B)$). The dependence
is obtained by quickly sweeping the bias voltage, while the
applied field is slowly changing. Fig.~\ref{fig2}-A displays
$r(V,B)$ when $B$ is OP. The cross in this image displays maximums
in resistance versus field and voltage, centered at zero field and
voltage, respectively.

The average resistance versus field and voltage are defined as
$r_0(B)=\int_{-V_{max}}^{V_{max}}r(V,B)dV/2V_{max}$ and
$r_0(V)=\int_{-B_{max}}^{B_{max}}r(V,B)dB/2B_{max}$, respectively,
where $B_{max}=12T$, and $V_{max}=4.2mV$. The averages are shown
in Figs.~\ref{fig2}-B and C. The resistance averaged over both $V$
and $B$ is $r_0=140\Omega$.

The average MR (Fig.~\ref{fig2}-B) at 0.03K is enhanced compared
to the AMR at 6K. This enhancement suggests that weak localization
effect contributes to MR at low temperatures. Prior research in
Co-films did not find any weak localization effects at
temperatures above $1.5K$.~\cite{brands} Our temperatures are much
lower than 1.5K, which could explain the difference between the
results.

The weak localization MR contribution cannot be extracted from
data because of the internal field of Co. The total magnetic field
acting on conduction electrons in Co cannot be equal to zero; it
is equal to the sum of the applied field and the internal field
(1.8T), which is much larger than the coercive fields. Thus, the
low-field contributions to quantum interference effects, such as
weak antilocalization, are experimentally inaccessible. In the
remainder of the text, we study CF only.

\begin{figure}
\includegraphics[width=0.45\textwidth]{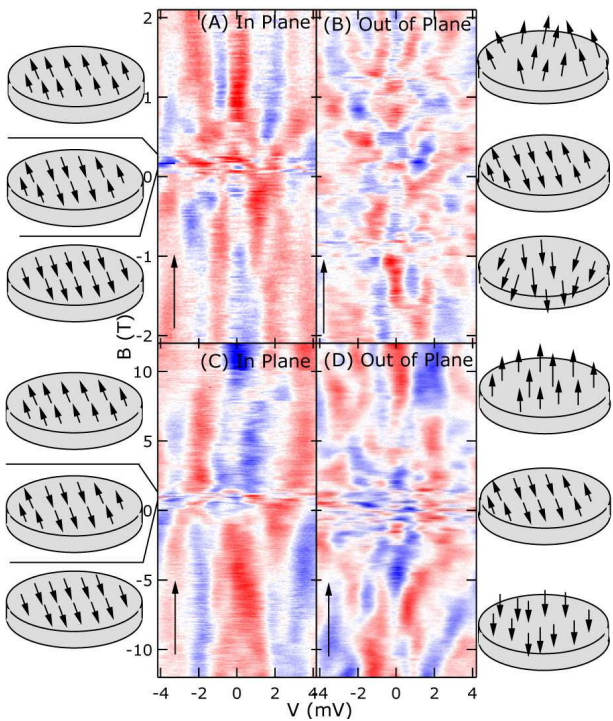}
\caption{A and B: Fluctuations in differential resistance,
$r(V,B)-r_0(V)-r_0(B)$, with $V$ and the IP-field and the
OP-field, respectively. C and D: same as A and B, but in a wider
field range. The minima and the maxima of differential resistance
with $B$ and $V$ correspond to constructive and destructive
electron interference, as described in text. The schematics
indicate the expected magnetic configurations.~\label{fig3}}
\end{figure}

The resistance maximum with voltage in Fig.~\ref{fig2}-C is a
consequence of the electron-electron interaction enhancement
effect (EE) in mesoscopic samples.~\cite{nagaev} Resistance
fluctuations in Fig.~\ref{fig2}-A are superimposed with the
EE-effect~\cite{nagaev} and the AMR. To better display the
fluctuations in resistance with voltage, we find the difference
between $r(V,B)$ and the average EE-effect in Fig.~\ref{fig2}-C.
The resulting resistance as a function of voltage at fixed field
is shown in Fig.~\ref{fig2}-D. The resistance now clearly exhibits
fluctuations with voltage. The fluctuations are reproducible and
are found at low temperatures only.

The fluctuations in $r$ with $V$ represent changes in electron
interference from constructive to destructive as a function of
electron energy.~\cite{washburn} Root-mean-square (rms) of the
fluctuations is $0.1\Omega$, which corresponds to rms-CF of
$0.05e^2/h$. The fluctuation amplitude is $\ll e^2/h$, showing
that the dephasing time must be much shorter than the transport
time. The rms does not change with $V$, showing that the heating
effects are weak.

The correlation voltage $V_C$ is given by the average spacing
between minima and maxima, $V_C\sim 0.5mV$. The meaning of $V_C$
is that changing the electron energy by $eV_C$ changes the
electron-phase by $\pi$. $V_C$ is related to the dephasing time
$\tau_\phi$ as $|e|V_C=\hbar/\tau_{\phi}$,~\cite{datta} so
$\tau_\phi =1.3ps$.

{\it Discussion of the results:} To display the fluctuating part
of the resistance, we subtract the average EE-effect and MR from
the resistance. Four red-white-blue images in Fig.~\ref{fig3}
display CFs with magnetic field and bias voltage at 0.03K
refrigerator temperature. Red and blue regions indicate larger and
smaller resistance, respectively.

The fluctuations with field and voltage, represented in
Fig.~\ref{fig3}, are not reproducible when the field is varied
arbitrarily. However, the fluctuations are reproducible when $B$
varies between two fields in the same direction after an initial
training with one field cycle.

In Figs.~\ref{fig3}-A and B, there is a noticeable difference
between dependence of the bias-fingerprints on the IP and the
OP-fields. If the IP-field varies from -2T to 0, the resistance
maxima and minima with voltage shift weakly. By contrast, when $B$
is changing OP, bias-fingerprints shift or rearrange several
times. These rearrangements are indicated by the appearances and
disappearances of the red and blue regions in Fig.~\ref{fig3}-B.

Fig.~\ref{fig3}-C shows that bias fingerprints vary weakly with
the IP-field in the field range $-12T<B<0$. But, when the IP-field
changes sign and reaches the coercive field, the bias-fingerprints
rearrange at the coercive field. This shows that the domain-walls
generate significant electron-phase shifts, at least on the order
of $\pi$.

Before discussing the physical origin of the rearrangements in
bias-fingerprints, we analyze the strong field data in
Figs.~\ref{fig3}-C and D, $2T<|B|<12T$. Comparing
Fig.~\ref{fig3}-C and Fig.~\ref{fig3}-D in this field-range, we
observe that bias fingerprints vary faster with the OP-field.
Alternatively, the characteristic field scale, which rearranges
the bias-fingerprint in strong field, is smaller in the
OP-direction. The correlation field $B_C$ is the average spacing
between the red and the blue regions along $B$ axes; $B_C\approx
4T$ and it is weakly dependent on $V$, confirming that the heating
is not significant.

In a strong OP-field, magnetization is saturated; and $B_C$ is
given by the field for a flux quantum over the phase coherent
area, $\Phi_0/L_{\phi}^2$, where $\Phi_0=h/e$ is the flux quantum,
and $L_{\phi}$ is the dephasing length. We find $L_{\phi}\approx
30nm$. Assuming a mean-free-path $l=5nm$ and the Fermi velocity
$v_F = 1.4\cdot 10^6 m/s$, the electron dephasing time is
$L_{\phi}^2/(v_F l/3)=0.4ps$, in agreement with $\tau_\phi$
obtained before, within an order of magnitude. In the
IP-direction, $B_C$ is larger; because the phase coherent area
perpendicular to the field is smaller,
$B_C=\Phi_0/tL_{\phi}\approx 12T$.

Now we discuss the rearrangements in bias-fingerprints at the
coercive fields, Fig.~\ref{fig3}-A and C. The internal field
switches at the coercive fields. In Co, the internal field change
is less than $3.6T$, much smaller than the IP-field for a flux
quantum ($12T$). So the Aharonov-Bohm effect cannot be responsible
for the rearrangements.

Fig.~\ref{fig3}-D shows that the density of red and blue regions
increases when $|B|<1.5T$. In Fig.~\ref{fig3}-B, there are about
five red and blue regions along B-axes between $0$ and $1.5$T.
This shows that the magnetization rotation from IP to OP direction
creates a phase-shift along a typical phase coherent electron
trajectory of about 5$\pi$. Since the total field (internal plus
applied) changes by $<3.3T$ in this applied field range and the
OP-field for a flux quantum is $4T$, five resistance minima and
maxima cannot originate from the Aharonov-Bohm effect.

We show that the phase-shift could originate from a weak
mistracking effect when conduction electron spins lag behind the
magnetic moments in the domain-wall.~\cite{greg,viret,levy,ebels}
The conduction electron spin tracks the local exchange field well
when the angular rotational period around the exchange field is
much smaller than the time of flight across the wall, which is
equivalent to a large value of the tracking parameter
$\xi=2E_{ex}\delta_w/hv_F$. Here, $E_{ex}$ is the exchange energy
between conduction electron spins and the spins responsible for
ferromagnetism. After a conduction electron traverses the wall,
the angle between the exchange field direction and the conduction
electron spin is $\theta=1/\xi$.

Using $E_{ex}=1eV$ and $\delta_w=15nm$, one obtains $\xi\approx
7.3$.~\cite{greg} The angular deviation increases the effective
potential energy of the conduction electron by $\Delta=E_{ex}
(1-cos(\theta))\approx 9meV$. The increase in effective potential
energy contributes to DWR.~\cite{greg,viret,levy,ebels} In
mesoscopic transport, however, electrons interfere among
trajectories with diffusion times shorter than $\tau_\phi$, which
leads to a correction in sample resistance. The wavefunction
attains a phase-shift from this effective potential of
$\Delta\tau_\phi/\hbar \approx 4.8\pi$. The phase-shift is reduced
to zero when the magnetic moments become parallel with each other.
So the bias-fingerprints should rearrange about 5 times when they
rotate into the OP-direction.

The dephasing length of $L_\phi=30nm$ is very short. In a separate
experiment, we measured two Co nanowires of lengths $500nm$ and
$800nm$ and width $100nm$ at $T=0.03K$(not shown). These nanowires
displayed no conductance fluctuations, confirming that $L_\phi\ll
500nm$.

The dephasing process in ferromagnets is not well understood. In
permalloy, experiments suggest that two level systems are
important sources of dephasing.~\cite{lees} Theoretically,
domain-walls were are found to reduce the dephasing
length.~\cite{tatara} But, the dephasing time $\tau_\phi
=\hbar/eV_C$ in our samples is independent of B, because $V_C$
does not vary significantly with B; $\tau_\phi$ in strong field,
in a single domain state, is approximately the same as $\tau_\phi$
at $B=0$ when domains are present. This demonstrates that the
domain-walls are not responsible for short $\tau_\phi$.

The phase of the wavefunction is extremely sensitive to the
position/presence of  domain-walls, as indicated by the
rearrangement of bias fingerprints in Fig.~\ref{fig3}. The absence
of domain-wall contribution to dephasing suggests that the
electron interaction with the wall must be elastic. This situation
is analogous to the sensitivity of conductance fluctuation with
respect to changes in the impurity
configurations.~\cite{washburn,lee2,altshuler4} In a thin film
mesoscopic sample, motion of an impurity by the Fermi wavelength
rearranges CFs. Nevertheless, the impurities do not contribute to
dephasing when electron scattering is elastic.

Kasai, {\it et al.}, found very short $L_\phi$ in Ni,
$L_\phi\approx 80nm$.~\cite{kasai} Small $L_\phi$ is correlated
with the large magnetocrystalline anisotropy in Ni; the dephasing
length in permalloy, which has negligible magnetocrystalline
anisotropy, is 500nm. Since the magnetocrystalline anisotropy in
Co is stronger than that in Ni, the dephasing length of 30nm in Co
agrees  with the trend that $L_\phi$ decreases with
magnetocrystalline anisotropy.~\cite{kasai}

 In conclusion, we
demonstrate mesoscopic resistance fluctuations induced by the
magnetization-reversal process in a Co nanoparticle. The
fluctuations are explained by the spin mistracking effect in
electron transport through domain-walls. The dephasing length at
low temperatures is only $30nm$, which is attributed to the large
magnetocrystalline anisotropy in Co, in agreement with the trend
established before, but not understood theoretically.

This work is performed in part at the Georgia Tech electron
microscopy facility. We thank P. Brouwer and A. Zangwill for
valuable discussions. his research is supported by the DOE grant
DE-FG02-06ER46281 and the David and Lucile Packard Foundation
grant No. 2000-13874.

\bibliography{career1}

\end{document}